\documentclass[aps,12pt,onecolumn,preprintnumbers,amsmath,amssymb,superscriptaddress]{revtex4}

\usepackage[english]{babel}
\usepackage{graphicx}
\usepackage{epstopdf}
\usepackage{color}
\usepackage{ragged2e}
\usepackage{float}
\usepackage{xcolor}
\usepackage{times}
\usepackage{pifont}
\usepackage[final]{pdfpages}
\begin{document}

\newcommand{\unit}[1]{\:\mathrm{#1}}         
\newcommand{\To}{\mathrm{T_0}}
\newcommand{\Tp}{\mathrm{T_+}}
\newcommand{\Tm}{\mathrm{T_-}}
\newcommand{\EST}{E_{\mathrm{ST}}}
\newcommand{\Rp}{\mathrm{R_{+}}}
\newcommand{\Rm}{\mathrm{R_{-}}}
\newcommand{\Rpp}{\mathrm{R_{++}}}
\newcommand{\Rmm}{\mathrm{R_{--}}}
\newcommand{\ddensity}[2]{\rho_{#1\,#2,#1\,#2}}
\newcommand{\ket}[1]{\left| #1 \right>}
\newcommand{\bra}[1]{\left< #1 \right|}
\newcommand{\braket}[2]{\left\langle #1\middle|#2\right\rangle}            
\newcommand{\ketbra}[2]{\left|#1\middle\rangle\middle\langle #2\right|}

\title{Signatures of collective photon emission and ferroelectric ordering of excitons near their Mott insulating state in a WSe$_2$/WS$_2$ heterobilayer}

\author{Luka Matej Devenica$^{\dagger}$}
\affiliation{Department of Physics, Emory University, 30322 Atlanta, Georgia, USA}

\author{Zach Hadjri$^{\dagger}$}
\affiliation{Department of Physics, Emory University, 30322 Atlanta, Georgia, USA}

\author{Jan Kumlin$^{\dagger}$}
\affiliation{Institute for Theoretical Physics, TU Wien, Wiedner Hauptstraße 8-10/136, A-1040 Vienna, Austria}

\author{Runtong Li}
\affiliation{Department of Physics, Emory University, 30322 Atlanta, Georgia, USA}

\author{Weijie Li}
\affiliation{Department of Physics, Emory University, 30322 Atlanta, Georgia, USA}

\author{Daniel Suarez Forrero}
\affiliation{Department of Quantum Matter Physics, University of Geneva, Geneva, Switzerland}

\author{Valeria Vento}
\affiliation{Department of Quantum Matter Physics, University of Geneva, Geneva, Switzerland}

\author{Nicolas Ubrig}
\affiliation{Department of Quantum Matter Physics, University of Geneva, Geneva, Switzerland}

\author{Song Liu}
\affiliation{Department of Mechanical Engineering, Columbia University, 10027 New York, New York, USA}
\affiliation{Institute of Microelectronics, Chinese Academy of Sciences, Beijing 100029, China}

\author{James Hone}
\affiliation{Department of Mechanical Engineering, Columbia University, 10027 New York, New York, USA}

\author{Kenji Watanabe}
\affiliation{Research Center for Functional Materials, National Institute for Materials Science, 1-1 Namiki, Tsukuba 305-0044, Japan}

\author{Takashi Taniguchi}
\affiliation{International Center for Materials Nanoarchitectonics, National Institute for Materials Science,  1-1 Namiki, Tsukuba 305-0044, Japan}

\author{Thomas Pohl$^\ddagger$}
\affiliation{Institute for Theoretical Physics, TU Wien, Wiedner Hauptstraße 8-10/136, A-1040 Vienna, Austria}

\author{Ajit Srivastava$^{\ast}$}
\affiliation{Department of Physics, Emory University, 30322 Atlanta, Georgia, USA}
\affiliation{Department of Quantum Matter Physics, University of Geneva, Geneva, Switzerland}

\maketitle
\justify
$^{\dagger}$These authors contributed equally to this work \\
$^{\ast}$Correspondence to: ajit.srivastava@emory.edu \\
$^{\ddagger}$Correspondence to: thomas.pohl@itp.tuwien.ac.at


{\bf 
Spontaneous symmetry breaking, arising from the competition of interactions and quantum fluctuations, is fundamental to understanding ordered electronic phases. Although electrically neutral, optical excitations like excitons can interact through their dipole moment, raising the possibility of optically active ordered phases. The effects of spontaneous ordering on optical properties remain largely unexplored. Recent observations of the excitonic Mott insulating state in semiconducting moir\'e crystals make them promising for addressing this question. Here, we present evidence for an in-plane ferroelectric phase of dipolar moir\'e excitons driven by strong exciton-exciton interactions. We discover a surprising speed-up of photon emission at late times and low densities in excitonic decay. This counterintuitive behavior is attributed to collective radiance, linked to the transition between disordered and symmetry-broken ferroelectric phases of moir\'e excitons. Our findings provide first evidence for strong dipolar inter-site interactions in moir\'e lattices, demonstrate collective photon emission as a probe for moir\'e quantum materials, and pave the way for exploring cooperative optical phenomena in strongly correlated systems.
}


Moir\'e crystals consisting of vertically stacked semiconducting transition metal dichalcogenides (STMDs) are being vigorously investigated for strongly correlated and topological electronic phases~\cite{cai2023signatures,park2023observation,zeng2023thermodynamic,XuNature2020,ReganNature2020,TangNature2020,ShimazakiNature2020,SmolenskiNature2021, ZhouNature2021}. Modulation of the electronic potential at the moir\'e length scale also affects excitons and the resulting moir\'e excitonic wavefunction has support primarily at the moir\'e sites~\cite{JinNature2019, TranNature2019, karni2022structure}. Thus, it is natural to ask whether, in addition to electronic phases, strongly correlated phases of lattice excitons can exist in moir\'e quantum materials. In fact, at unit filling, onsite interactions between excitons in a lattice can result in a correlated excitonic state that resembles a Mott insulator. The recent observation of such a phase has opened the field for the study of Bose-Hubbard physics of excitons~\cite{lagoin2022mott,xiong2023correlated,park2023dipole, wang2023intercell, gao2024excitonic, lian2024valley}. 

In a typical type-II semiconducting heterostructure, electron and hole are localized in separate layers leading to long-lived, spatially indirect interlayer excitons (IXs). IXs feature an out-of-plane static dipole moment making them good candidates for realizing phases arising from finite-range interactions~\cite{rivera2016valley,liNMat2020dipolar,kremserNpj2D2020discrete, lagoin2022extended, lagoin2024superlattice}. The addition of the moir\'e potential can give rise to a richer internal structure of moir\'e excitons that results in qualitatively different excitonic properties and interactions. 
The moir\'e unit cells, which may be thought of as `moir\'e atoms', can contain multiple minima for electrons and holes, allowing for different `moir\'e orbitals'. For example, electrons and holes in different moir\'e orbitals can have an in-plane separation that greatly exceeds the out-of-plane displacement set by the layer separation, resulting in a large static dipole moment, and consequently strong exciton-exciton interactions. Moreover, excitons are naturally dissipative towards photon emission, which we use as a probe of ordered phases, and explore the possibility of modifying collective emission through exciton-exciton interactions, which opens the door to studying cooperative optical phenomena in quantum materials~\cite{kumlin2024superradiancestronglyinteractingdipolar, huang2024collective, lagoin2024evidence}. 

Here, we provide evidence for an in-plane ferroelectric ordering of static dipoles of moir\'e excitons in an H-stacked WSe$_2$/WS$_2$ heterobilayer close to the excitonic Mott state. Whereas the onsite hole-interaction converts in-plane quadrupolar excitons to dipolar excitons above a certain filling, their finite-range dipole-dipole interactions close to unit filling align the dipoles in an energetically favorable head-to-tail configuration. Given the different moir\'e orbitals discussed above, the excitonic quantum state has a three-fold degeneracy with respect to the orientation of the in-plane dipole, corresponding to a discrete rotational symmetry. The full hybridization of these states implies a vanishing net static dipole moment and also results in a threefold collective enhancement of the excitonic photon emission rate. In contrast to such quadrupolar excitons, the symmetry-broken state with a maximal in-plane dipole consequently features a three times longer radiative lifetime. 
Unlike in atomic systems, the hybridization that causes collectively light emission is not mediated by the electromagnetic vacuum, but arises from the tunnel coupling of the hole across the three configurations. 

Figure 1a shows the relative position of the electron and hole in a moir\'e unit cell of the lattice-mismatched WSe$_2$/WS$_2$ heterostructure, as determined by DFT calculations in previous works~\cite{wang2023intercell, lian2024valley}. In the R-stacked (0$^\circ$ stacking angle) case, the potential minima for the hole and electron have no lateral displacement between them, such that the exciton has a static dipole moment in the out-of-plane direction, but no in-plane dipole or higher-order moment. In contrast, the H-stacked (60$^\circ$ stacking angle) case features laterally displaced potential minima of the electron and hole. For a given position of the hole (electron) of the exciton, the electron (hole) is free to delocalize over the three equivalent neighboring sites resulting in an in-plane quadrupolar moment, as shown in Fig.~1a. In this work, we will primarily focus on H-stacking to explore the resulting in-plane quadrupolar and dipolar nature of the moir\'e excitons.

In order to confirm the presence of the moir\'e superlattice in our sample, we first studied photon emission as a function of carrier doping. Figure 1b shows the photoluminescence (PL) emission from IX in the H-stacked region, as a function of electron doping in the sample. Our dual-gated sample geometry (Supplementary Fig.~S1) permits independent control of the doping and the out-of-plane displacement field. Consistent with  previous observations~\cite{park2023dipole, xiong2023correlated, wang2023intercell, gao2024excitonic, lian2024valley}, we measure a blueshifted peak in PL (IX$_\mathrm{e}$) at unit electron filling, which arises from the onsite exciton-electron repulsion U$_{\mathrm{Xe}}$ $\simeq$ 42 meV. In addition, we observe a characteristic modulation of IX PL at other fractional fillings from $1/3$ to unit filling, consistent with the formation of generalized Wigner crystals of electrons at fractional fillings~\cite{ReganNature2020, TangNature2020, li2021imaging}. An interesting exception to this pattern occurs at 1/7 filling, where we observe a sudden and sizeable ($\simeq$ 13 meV) redshift of the emission. As in previous work~\cite{wang2023intercell}, we interpret this as the formation of an intercell exciton complex between an exciton and an electron in neighboring moir\'e sites. The quadrupolar IX becomes polarized because of an attractive interaction with a neighboring electron, and the 1/7 electron Wigner crystal guarantees that each exciton will have at least one neighboring IX (Supplementary Fig.~S2). Indeed, at 1/7 electron doping, all IXs are redshifted. Therefore, we label the state at 1/7 filling as the intercell IX-electron complex (IX$_{\mathrm{ICe}}$). Importantly, this redshift could not be observed in previous studies on R-stacking, where electrons and holes have no in-plane displacement. This difference provides strong evidence for the in-plane quadrupolar nature of moir\'e excitons in H-stacked bilayers~\cite{wang2023intercell, lian2024valley}. Moreover, the measurement of the IX g-factor of 10.5 $\pm$ 0.5 (Supplementary Fig.~S3) confirms our assignment of the H-stacked region~\cite{TangNature2020, montblanch2021confinement}.

Having established the presence of a moir\'e lattice in our sample, we study its photon emission properties. Using pulsed laser excitation, we can measure the dependence of the PL spectrum on the excitation power (see Methods). As seen in Fig.~1c, at a certain power threshold, another blueshifted peak appears. In accordance with recent works~\cite{park2023dipole,xiong2023correlated,lian2024valley,gao2024excitonic}, we attribute this peak (IXX) to two excitons sharing a single moir\'e site, which occurs at a sufficiently high exciton density. 
We further confirm this by observing IXX's distinctive lack of circular polarization~\cite{wang2023intercell} (Supplementary Fig.~S4). By comparing spectra at low and high excitation power, we deduce an on-site repulsion energy of $\approx$ 36 meV, which makes the state at exactly one-exciton-per-moir\'e site filling an excitonic Mott insulator. The integrated total emission intensities of IX and IXX further confirm the Mott state (Fig.~1d). The IX intensity increases linearly with the excitation until we observe the additional IXX peak, and the IX intensity begins to saturate. We label the power at which we begin to observe IXX emission as $\nu_{\mathrm{max}} = 1$ (Supplementary Note 1). Beyond $\nu_{\mathrm{max}} = 1$, the IXX peak grows linearly, as shown in the inset of Fig.~1d where the power axis was shifted by its value at  $\nu_{\mathrm{max}} = 1$. This indicates that IXX appears only when all the available nearby IX sites are filled (Supplementary Figs.~S5 and S6). Note that, the linear power dependence of the IX intensity below $\nu_{\mathrm{max}} = 1$ permits a direct conversion between excitation power and exciton filling  $\nu_{\rm max}$  (Supplementary Note 1).


In order to explore potential effects of exciton-exciton interactions in the moir\'e lattice, we now study the density dependence of the excitonic properties. To this end, we have recorded the power dependence of PL spectrum under continuous-wave (CW) driving, which establishes a steady-state exciton density  for a given power. As shown in Fig.~2a, we observe a red-shifted peak to emerge at an excitation  power below the critical power for creating IXX. With increasing pump power, this red-shifted exciton signal, which we denote as IX$_\mathrm{d}$, rapidly exceeds the low-density IX-peak and eventually dominates the entire exciton emission at high densities. Importantly, we do not observe the red-shifted IX$_\mathrm{d}$ peak in R-stacked regions of the sample under identical CW excitation, as seen by comparing Figs.~2c and 2d. The opposite slope of the Stark shift with the displacement field in the R- and H-stacked regions comes from our sample geometry (Supplementary Fig.~S1), and confirms our assignment of the measured regions (Supplementary Fig.~S7). This indicates that the emerging red-shifted IX$_\mathrm{d}$ peak originates from the excitonic in-plane dipole, which is only present in the H-stacked case. In fact, the coexistence of IX and IX$_\mathrm{d}$, together with the absence of a gradual energy shift with increasing powers, rules out power-dependent sample heating as a cause for the emergence of the IX$_\mathrm{d}$ peak. Below, we present further evidence that this surprising density-dependence of the exciton emission indeed arises from exciton-exciton interactions due to their in-plane dipoles in H-stacked bilayers.


To this end, we have measured the decay dynamics of the excitons after pulsed excitation (see Methods).  Figures~3a and 3b show the time evolution of the total PL signal from the combined IX and IX$_\mathrm{d}$ peaks for different initial lattice filling fractions $\nu_{\rm max}$, controlled by the power $P_{\rm exc}$ of the excitation pulse. Up to an initial excitonic filling of $\nu_{\mathrm{max}} \approx 0.36$, we observe a simple exponential decay  with a lifetime of $160 \pm 10$ ns. As the filling increases beyond this value, the initial decay dynamics slows down steadily (Fig.~3b), and the initial decay time saturates to a three times larger value of $460 \pm 10$ ns near the excitonic Mott state $\nu_{\mathrm{max}} = 1$. We rule out unintentional photodoping as a reason for this slowdown (Supplementary Note 3, Supplementary Fig.~8).
At longer times, the excitonic decay speeds up again with a decay time that corresponds to the value found at lower initial densities. A further increase of the excitation power has no notable effect on the decay dynamics, which resembles the unit-filling behavior, as shown in the inset of Fig.~3b. This indicates that excitons in doubly occupied sites (IXX) generated at high excitation power undergo rapid decay, leaving behind a near-unit-filled lattice. In fact, this process explains the small non-monotonic `burst' in the first 30 ns of the high-power data (Supplementary Note 2, Supplementary Fig.~S6). The observed decay dynamics, indeed, comes as a surprise, as an increased exciton density often accelerates decay because of non-radiative processes~\cite{miller2017long, JaureguiScience2019} and fast decay processes typically tend to first deplete the rapidly decaying components, generically leading to fast decay followed by a slow evolution \cite{choi2021twist,MoodyJOSAB2016,zhang2015experimental}.

For a more detailed analysis, we fit the data in Fig.~3b by two exponentials and define the transition time, $t_{\rm tr}$, as the crossing of the two curves Fig.~3c. By integrating the decay curve before and after this point, we are able to determine the filling fraction, $\nu_{\rm tr}$, of excitons that remains at the transition time. Figure 3d shows $\nu_\mathrm{tr}$ as a function of the initial filling $\nu_{\mathrm{max}}$, and demonstrates that the exciton decay rate changes dynamically at the same filling fraction of $\approx 0.27 \pm 0.04$. In Fig.~3d, we also show the initial decay lifetime as a function of $\nu_{\rm max}$. As discussed above, the initial decay time remains independent of filling until a critical value $\nu_{\rm cr}$ followed by a gradual increase to a three-fold value at $\nu_{\rm max}=1$. Remarkably, this critical value $\nu_{\rm cr}\approx\nu_{\rm tr}$ coincides with the filling fraction at which the dynamical transition between the two decay regimes occurs. This provides strong experimental evidence that the observed decay characteristics is indeed related to the exciton density, causing a slowdown of photon emission above a lattice filling of $\sim1/3$.

In order to understand this puzzling slow-to-fast transition of the decay, we first focus on the filling fraction $\nu_\mathrm{tr}\approx 1/3$, which marks the transition between faster asymptotic decay and slower initial decay. As depicted in Fig.~3e, for fillings above $\nu_\mathrm{max}$ = 1/3, there would be a guaranteed carrier-carrier overlap if all excitons were to remain quadrupolar. For two adjacent excitons the strong onsite Coulomb repulsion between holes ($\sim$ 50 meV) \cite{WuPRL2018, TangNature2020} greatly exceeds the energy gain from hybridizing the three states of each exciton. This implies a constraint on hole delocalization leading to a destabilization of the quadrupolar IX in favor of a polarized dipolar IX, as shown in Fig.~3e. Therefore, for $\nu_\mathrm{max} > 1/3$, the ratio of dipolar to quadrupolar excitons increases. Indeed, this critical density is close to the observed values of $\nu_{\rm tr}$ and $\nu_{\rm c}$ for the transition between slow and fast decay.

Let us now consider the effect of such a polarization of the quadrupolar exciton on its radiative decay rate. For a given electron position in the lattice, an isolated moir\'e exciton ($\nu_{\mathrm{X}} < 1/3$) can bind in one of three degenerate states, corresponding to different orientations of the excitonic dipole (static) and which we denote as $\lvert1\rangle$, $\lvert2\rangle$, and $\lvert3\rangle$, as shown in Fig.~3f. The rate, $\gamma_{\rm d}$ of photon emission from each of these polarized dipolar states is proportional to the square of the transition dipole moment ($\hat{d}$)
\begin{equation}
\gamma_{\rm d}\propto \lvert\langle i\rvert\hat{d}\lvert 0\rangle\rvert^2\quad,\:i=1,2,3
\end{equation}
where $\lvert0\rangle$ denotes the vacuum state at a given site. Owing to their threefold rotational symmetry, we expect isolated moir\'e excitons to form a hybridized superposition state $\lvert Q\rangle = \frac{1}{\sqrt{3}}\left( \lvert 1\rangle + \lvert 2\rangle +\lvert 3\rangle \right)$. In fact, this quadrupolar exciton can be considered as a few-body superradiant state, which -- due to its deep sub-wavelength spatial extent -- features a collectively enhanced photon emission rate 
\begin{equation}
\gamma_Q \propto \lvert\langle Q \rvert \hat{d}\lvert 0\rangle\rvert^2 = \left|\frac{\langle 1\rvert + \langle 2\rvert +\langle 3\rvert}{\sqrt{3}} \hat{d}| 0\rangle\right|^2 = 3\gamma_d.
\end{equation}
Note that the transition dipole moment ($\hat{d}$) is unrelated to the static dipole, such that the polarization of emitted photons does not depend on the orientational state ($\lvert i\rangle$) of the static dipole.
Hence, we conclude that the fully hybridized quadrupolar exciton  should decay three times faster than a polarized dipolar exciton.

We have performed microscopic simulations of the correlated decay dynamics using a Monte Carlo approach (Supplementary Note 4). For a given spatial configuration of excitons in the lattice, we perform a Monte Carlo sampling of their dipoles, including the orientational constraint due to hole repulsion. This yields corresponding decay rates which are used to propagate the system via Monte Carlo sampling of stochastic quantum jumps. Offering full information about the time evolving many-body state of the lattice, this approach permits to calculate the dynamics as well as spectral properties of photon emission. Figure~3g shows a comparison of the simulation results and the measured signal for different initial lattice fillings (see Supplementary Note 4 and Note 5, Fig.~S9 for details). We find fairly good agreement between the experiment and simulations, in particular considering the simplicity of the theoretical description. Importantly, the theory predicts the transition to a collectively enhanced decay at the lattice filling of $1/3$, as observed experimentally. At high densities and short times, the simulations yield a non-exponential decay dynamics that is not observed experimentally, which may be attributed to additional decoherence processes and additional weak orientational constraints due to finite-range dipole-dipole interactions that are not accounted for in the simulations.

The developed theoretical picture suggests a close correlation between the two-stage decay dynamics and the nonlinear spectral response of moir\'e excitons in our sample. We have explored this question further by measuring time-resolved PL spectra (TRPLS) with an electronic streak camera technique (see Methods and Supplementary Fig.~S10) in order to compare the PL emission energies from the IX and IX$_\mathrm{d}$ peaks in the two lifetime regimes. Spectrally binning the data in the same range used for the lifetime measurements confirms that the observed TRPLS decay is consistent with lifetime measurements discussed above (Supplementary Fig.~S11). Figure 4a shows the TRPLS at low excitation power, where we observe exponential decay with a single rate $\sim3\gamma_d$. At high excitation powers, however, (see Fig.~4b) the emission starts out on the red-shifted IX$_\mathrm{d}$ peak, whose energy shift of $15$ meV coincides with that observed in the CW experiments, shown in Fig.~2a-b. As excitons decay and their density decreases, the emission energy returns to that of the IX peak, seen at low initial densities in Fig.~4a. Figure 4c shows the dynamics of the average emission energy together with the total PL signal. The measurement demonstrates that the energy transitions gradually from that of the dipolar IX$_\mathrm{d}$ exciton and reaches its maximum value corresponding to the quadrupolar IX during the slow initial regime of photon emission. The observed time evolution is, therefore, consistent with an initially ordered ferroelectric phase at high densities that features a lowered energy due to the in-plane interactions between aligned dipoles (Fig.~4d), and whose energy consequently rises as excitons decay until a filling of $\sim1/3$ where the system is primarily composed of isolated quadrupolar IX excitons (Fig.~4d). After transitioning to IX emission, a subsequent smaller 3 meV redshift is observed, which also occurs in the low-power case (Fig.~4a-c), and most likely arises from decreasing out-of-plane dipolar repulsion with decreasing densities~\cite{park2023dipole, lian2024valley}. This conclusion is further supported by the fact that the repulsive dipolar shift is also observed for R-stacking (Fig.~2d). 

For a quantitative analysis we can use our Monte Carlo simulations to calculate the expected energy dependence of our TRPLS data. Having access to the microscopic exciton configurations during the simulated decay dynamics, one can determine the energy of emitted photons from the dipole-dipole interaction energy at every Monte Carlo step. Including a Gaussian linewidth of $\Delta E = 10 \, \mathrm{meV}$ to account for typical broadening effects and averaging over many Monte-Carlo runs, we obtain the time-evolving emission spectra as shown in Fig.~4e-f. Indeed, the calculations capture the essential features of our TRPLS measurements. In particular, experiment and theory show the characteristic change of the emission spectrum due to the formation of a ferroelectric phase of ordered dipoles at high densities. 
We find that an in-plane dipole moment corresponding to $\sim$70\% of the distance between the single-particle potential minima of the electron and hole reproduces the observed $\simeq$ 15 meV redshift in the fully polarized ferroelectric phase. Owing to electron-hole attraction, such a reduction in the excitonic electron-hole separation appears reasonable since the moir\'e-potential depth is comparable to the exciton binding energy~\cite{shabani2021deep, wilson2017determination}. 

Our observations, thus, reveal a rich interplay between onsite Coulomb interactions, finite range dipole-dipole interactions, and collective photon coupling of excitons to emerge from the presence of in-plane excitonic dipole moments in moir\'e quantum materials. The physics of ferroelectricity in two-dimensional materials has recently attracted substantial interest, motivated by broad scientific and technological perspectives \cite{Seidel2023}. The demonstrated dynamical transition to ferroelectric ordering could be combined with an in-plane electric field to tune the spontaneous macroscopic polarization in an optically active quantum material. Our findings suggest new scientific venues -- from studying emergent ordered phases in dipolar quantum matter in and out of equilibrium ~\cite{de_paz_nonequilibrium_2013,zhang_equilibrium_2015,huber_nonequilibrium_2020,su_dipolar_2023} to exploring cooperative optics in ordered strongly interacting systems~\cite{rui_subradiant_2020, kumlin2024superradiancestronglyinteractingdipolar}. More specifically, our findings may enable further investigations into non-equilibrium physics of the quantum rotor or three-state Potts model~\cite{wu_potts_1982, park_three-state_1994},  the emission of non-classical light from strongly correlated states of quantum emitters \cite{de_vega_detection_2008, pizzi_light_2023} or  the probing and sensing of phases and phase transitions in quantum materials. 

\justify
\textbf{Methods}
\\
\textbf{Device fabrication} \\
We use a dry transfer fabrication method with a polycarbonate (PC) stamp to fabricate the dual-gated, hBN-encapsulated transition metal dichalcogenide devices~\cite{zomer2014fast}. WSe$_2$ and WS$_2$ monolayers, few-layer graphene flakes for gates and contact and hBN for encapsulation are mechanically exfoliated from bulk crystals onto 300 nm SiO$_2$ on Si substrates.  The thickness of the hBN flakes, estimated from optical contrast, ranges from 20 to 40 nm. Firstly, we use a PC stamp to pick up the bottom hBN and the bottom few-layer graphene, which is then deposited onto a 300 nm SiO$_2$/Si substrate with EBL-made electrodes (5 nm Cr/85 nm Au) at 170 $^\circ$C. The substrate with bottom hBN and graphene is then annealed in an 5\%H$_2$/95\%N$_2$ environment at 350 $^\circ$C for 3 hours to remove the PC residue. The few-layer graphene top gate, top hBN flake, top WS$_2$ monolayer, WSe$_2$ monolayer, bottom WS$_2$ monolayer, and the few-layer-graphene contact are picked up in that order with a PC stamp and then placed onto the back gate at 170 $^\circ$C. We use the tear-and-stack method~\cite{kim2016van} to pick up half of a large WS$_2$ flake as the top monolayer, align with the WSe$_2$ layer within 1$^\circ$ uncertainty, and pick up the other half of the WS$_2$ flake with a 60$^\circ$ rotation relative to the top WS$_2$, and offset the flakes laterally in order to create distinct R and H stacked regions, separated by a trilayer region (Supplementary Fig.~S1).

\justify
\textbf{Optical measurements and electrostatic gating} \\
Photoluminescence spectroscopy and emission lifetime measurements were performed inside two cryostats (AttoDry 800 and Bluefors BF-LD 4K), each featuring a homemade confocal optical microscope. All main text data was collected in the AttoDry 800 system, with the exception of Figure 1b, which was collected in Bluefors BF-LD 4K. In each cryostat, we use a piezoelectric controller (Attocube systems) to move the sample. The excitation laser used was Picoquant LDH-D-C-640, with a center wavelength of 640 nm and operated in pulsed (up to 1MHz) or continuous-wave mode. In both setups, an achromatic objective (NA = 0.42 for AttoDry 800 and  NA = 0.63 for Bluefors BF-LD 4K) is used to focus the laser beams to a spot size of $\sim$1 $\mu$m, which is then collected through the same objective and coupled into an optical fiber. The light was then directed either to a spectrograph or avalanche photodiode for emission lifetime measurements.

For spectrophotometry, the light was outcoupled into a high-resolution spectrograph (Princeton Instrument HR-500, focal length 500mm) for AttoDry 800 and Princeton Instruments SP-2750i (focal length 750mm) for BlueFors BF-LD 4K) which were both operated with a 300 grooves per mm grating (blazed at 750 nm). A charge coupled device (Princeton Instrument PIXIS-400 CCD for AttoDry 800 and PyLoN CCD for BlueFors) was used as the detector. The excitation and emission arms also each featured a liquid crystal retarder (Thorlabs LCC1513-B), which enabled us to excite the sample (and collect emission of light) with well-defined linear or circular polarization.

For emission lifetime measurements, the light was out-coupled to an avalanche photo-diode (PerkinElmer SPCM-AQR-16-FC), which was plugged into a photon counter (Picoquant PicoHarp 300). The sample emission was filtered with a combination of 780LP and 900(20)BP filters in order to simultaneously collect IX and IX$_\mathrm{d}$ and exclude IXX.

For the time-resolved photoluminescence spectra measurements (TRPLS), we used a fiber-coupled acousto-optic modulator (AOM) (Aerodiode 850-AOM-2) in-line with the fiber collection, which modulated the collection time window, after which the same setup as described in the previous paragraph collected the spectra (Supplementary Fig.~S10). The AOM ON-time was controlled by a tunable-width digital delay/pulse generator (SRS DG535), whose other channel also served as the trigger for the pulsed laser, ensuring synchronization across excitation and collection. The AOM ON-time was set to 10ns by sending 10ns-wide pulses from the SRS DG535 and operating the AOM controller in external trigger mode. The delay of the AOM ON-time pulse was then scanned, giving binned PL measurements over 10 ns intervals at arbitrary delays.

Through two source meters (Keithley 2400), we apply independent voltages to the graphene top gate and graphene bottom gate to tune the charge density or apply an electric field to the sample, while the contact to the sample is grounded to a ground common to the source meters.

\justify
\textbf{Emission Lifetime and Energy Simulation} \\
The emission lifetime and emission spectra simulations were performed using a Monte Carlo sampling and quantum jump approach on a triangular lattice with $30 \times 30$ sites under periodic boundary conditions (Supplementary Note 4). Excitons were treated as dipolar particles occupying distinct lattice sites, initialized in one of three possible hole configurations. For filling fractions below $\nu = 1/3$, we avoid the occupation of nearest-neighbor sites due to the dipolar repulsion and the fact that there is no slowdown in the decay measured below these filling fractions. Emission rates were determined by local exciton densities, with recombination probabilities calculated dynamically. For example, excitons with no occupied neighboring sites decayed with a rate of $\gamma_{\rm Q} = 6.3 \, \mathrm{MHz}$ (lifetime $\tau_{\rm Q} = 160 \, \mathrm{ns}$), while those with fully occupied neighboring sites had a reduced rate $\gamma_{\rm d} = \gamma_{\rm Q} / 3 = 2.1 \, \mathrm{MHz}$ (lifetime $\tau_{\rm d} = 480 \, \mathrm{ns}$), matching experimentally obtained high- and low-density lifetimes. Intermediate rates were assigned proportionally based on available hole states. The evolution followed a Lindblad master equation with dissipative recombination dynamics simulated through stochastic quantum jumps. Emission spectra were obtained by tracking energy shifts during the exciton decay, incorporating only nearest-neighbor Coulomb interactions. Quantitative agreement with the experimental data is obtained by reducing the in-plane distance of electrons and holes by a factor of 0.7 compared to their single-particle potential minima. 

\justify
{\bf Acknowledgments}
We thank Lida Zhang, M. Hafezi, A. Imamo\u{g}lu, Martin Claassen, Tomasz Smole\'nski, and Martin Kroner for insightful discussions. A.~S. acknowledges funding from the NSF Division of Materials Research (award No.~1905809), from the State Secretariat for Education, Research and Innovation (SERI)-funded European Research Council Consolidator Grant TuneInt2Quantum (No.~101043957). T.~P. and J.~K. acknowledge support from European Union’s Horizon Europe research and innovation programme under the Marie Sklodowska-Curie
grant agreement No.~101106552 (QuLowD), from the
Austrian Science Fund (Grant No.~10.55776/COE1) and
the European Union (NextGenerationEU), and from the European Research Council through the
ERC Synergy Grant SuperWave (Grant No.~101071882). Synthesis of WSe$_2$ (S.L., J.H.) was supported by the NSF MRSEC program through the Columbia University Center for Precision-Assembled Quantum Materials (DMR-2011738). K.~W. and T.~T. acknowledge support from the Elemental Strategy Initiative conducted by the MEXT, Japan (Grant Number JPMXP0112101001) and  JSPS KAKENHI (Grant Numbers 19H05790, 20H00354 and 21H05233).
\justify
{\bf Author contributions}

A. S., T.P., L. D., Z. H., W. L. conceived the project. K. W., T. T. provided the hBN crystal and S. L. and J. H. provided the WSe$_2$ crystals. L. D., W. L., Z. H. prepared the samples. L. D., Z. H., R. L., D. S. F., V. V., N. U. carried out the measurements. J. K. and T.P developed the theoretical model and conducted the Monte Carlo simulations. A. S., T. P. supervised the project. All authors were involved in analysis of the experimental data and contributed extensively.

\newpage

\begin{figure}
\includegraphics[width = 6in]{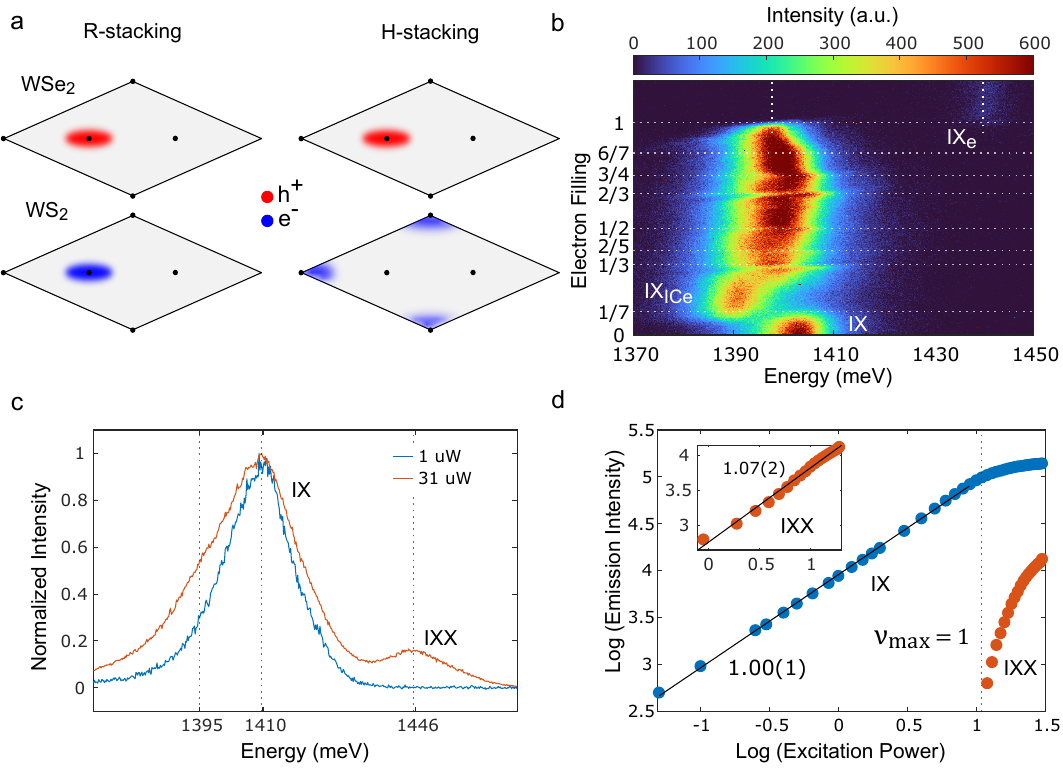}
\end{figure}

\noindent {\bf Figure 1: Interlayer Excitons in a WSe$_2$/WS$_2$ bilayer and their Mott insulating state.} 
{\bf a,} Diagram of electron and hole density in an R and H stacked WSe$_2$/WS$_2$ heterobilayer moir\'e unit cell. The red (blue) areas indicate regions of high hole (electron) density. The black line indicates the edge of a moir\'e unit cell, and the vertices indicate the high-symmetry registry points. Unlike in R-stacking, an exciton in H-stacking features a static in-plane quadrupolar moment.
{\bf b,} Doping scan of the photoluminescence (PL) emission of an H-stacked region, using 10 nW of 1 MHz 640 nm pulsed laser excitation. Dashed lines indicate generalized Wigner crystals at fractional fillings.
{\bf c,} PL emission of an H-stacked region at a high and low excitation power, using a 1 MHz 640nm pulsed laser. The difference in IXX to IX energy shows U$_{\mathrm{IXX}}$ = 36 meV. At higher power, a red shoulder at 1395 meV becomes prominent. 
{\bf d,} Log-log plot of total intensity of the IX and IXX peaks as a function of excitation power. The intensity of IX increases linearly (slope 1.00 $\pm$ 0.01) until the power at which IXX becomes measurable, when IX starts to saturate. The inset shows the IXX intensity against power, with an offset to the IX saturation power, which shows IXX intensity also increases linearly (slope 1.07 $\pm$ 0.02)) when the power is above the saturation value.

\newpage

\begin{figure}
\includegraphics[width = 6in]{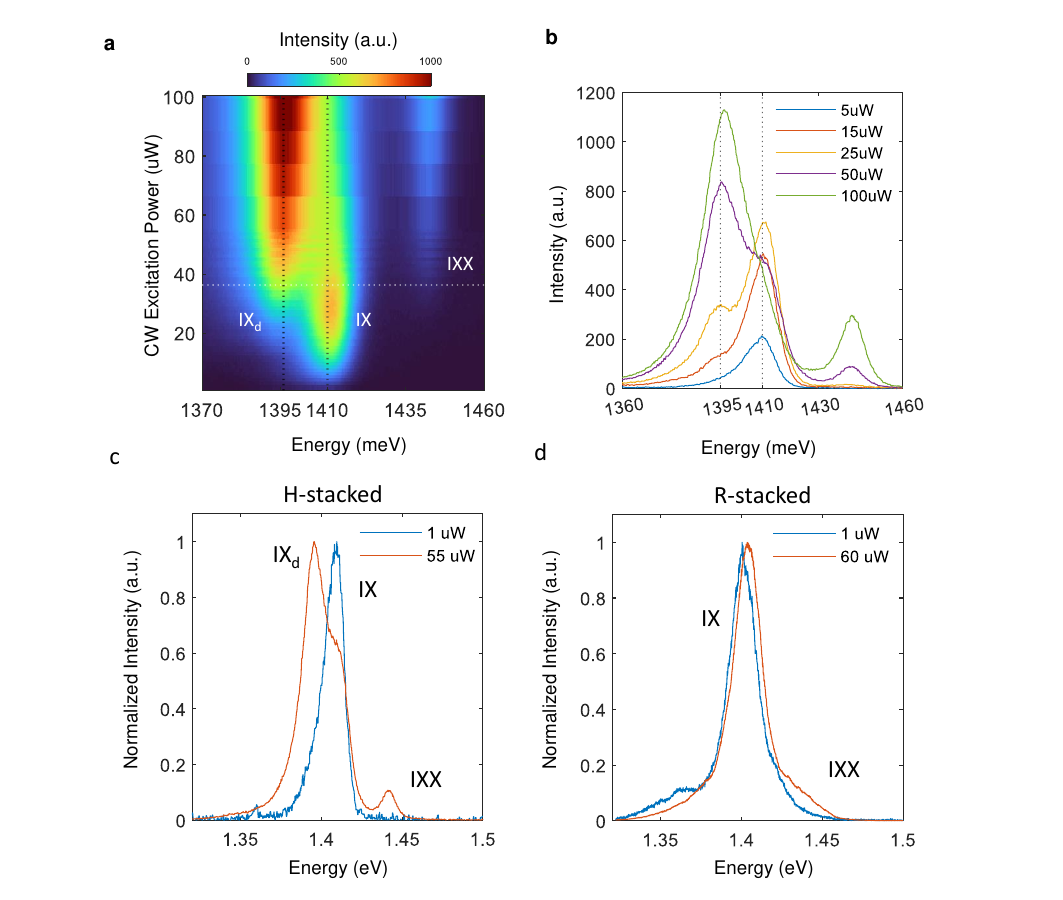}
\end{figure}

\noindent {\bf Figure 2: Red-shifted IX emission at high excitonic density.}
{\bf a,} IX PL as a function of excitation (640nm CW laser) power at charge neutrality. The horizontal dashed line indicates the power at which IXX appears.
{\bf b,} Select spectra from panel {\bf a} showing an IX$_\mathrm{d}$ red-shift of 15 meV. The IX$_\mathrm{d}$ peak appears before IXX and increases with power, at the expense of the IX peak.
{\bf c,} Normalized IX PL spectra at a low and high (640nm CW laser) power, in the H-stacked region of the sample.
{\bf d,} Normalized IX PL spectra at a low and high (640nm CW laser) power, in the R-stacked region of the sample.

\newpage

\begin{figure}
\includegraphics[width = 6in]{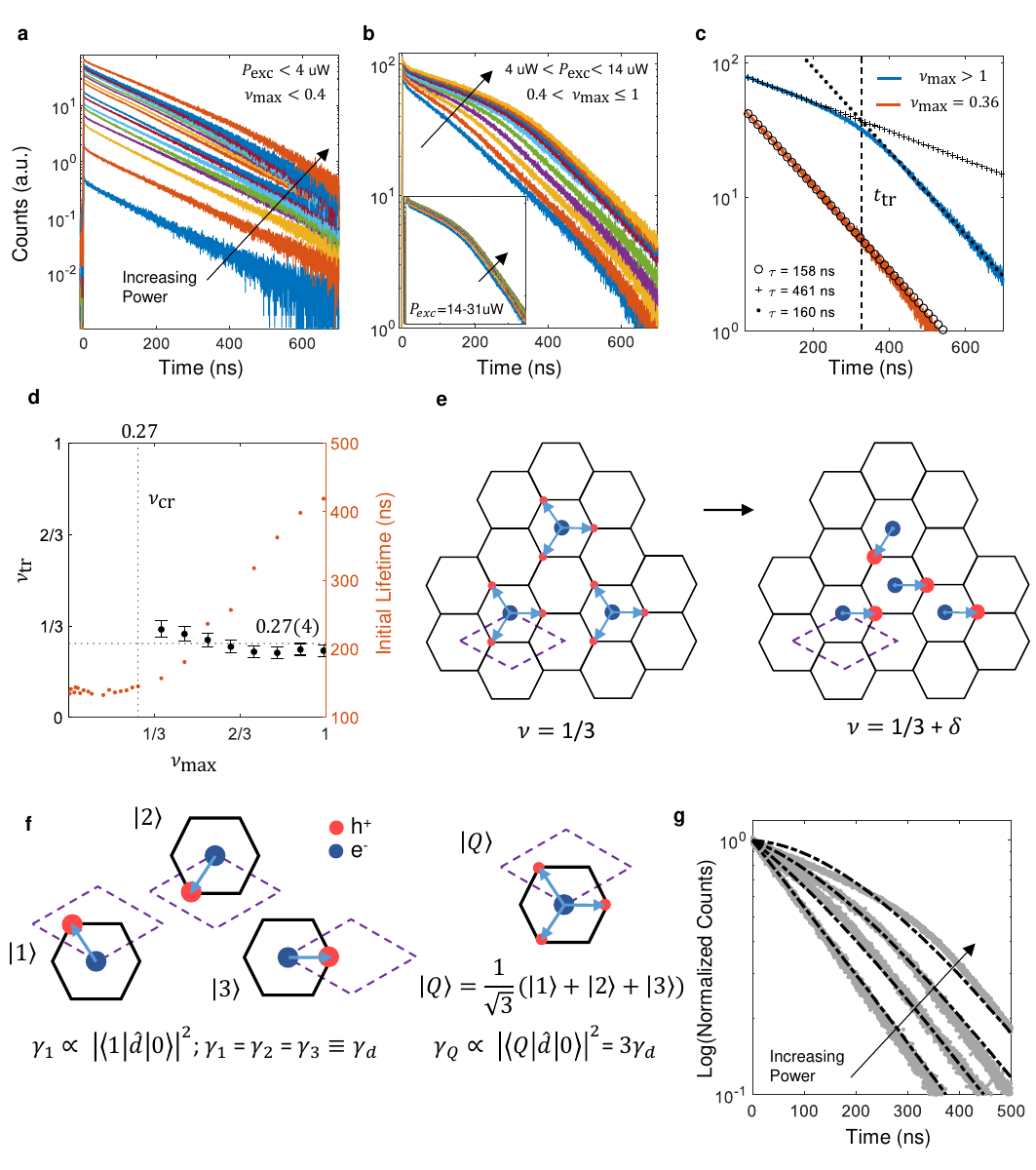}
\end{figure}

\noindent {\bf Figure 3: Temporally and spectrally resolved IX emission in the H-stacked region.}
{\bf a.} IX emission decay for the low excitation power regime ($\nu_\mathrm{max} <$ 0.4). The decay is exponential and lifetime is constant over the entire power range. The power indicated in panels {\bf a-c} of this figure is the average excitation power, using a 640 nm, 1 MHz pulsed laser. The collected emission in panels {\bf a-c} of this figure is filtered to only include IX and IX$_\mathrm{d}$ (see Methods).
{\bf b.} IX emission decay for the intermediate excitation power regime ((0.4 $< \nu_\mathrm{max} <$ 1)). The decay becomes initially slower, then returns to the low-density decay rate. Inset shows IX emission decay for the high excitation power regime ($\nu_\mathrm{max} >$ 1). The slowing effect has saturated beyond the excitonic Mott state. Axis scaling and limits on the inset are same as in parent figure.
{\bf c.}  IX emission decay of a high (blue) and low (orange) excitation density case. The fits show the slow and fast decay rates occurring at the high and low density case, respectively. Dashed line indicates the time of transition $t_\mathrm{tr}$ from the slow to the fast decay regime, corresponding to a density $\nu_\mathrm{tr}$.
{\bf d.} Left axis (black) shows the transition fraction $\nu_\mathrm{tr}$ as a function of the inital density $\nu_\mathrm{max}$. Dashed line indicates the mean observed $\nu_\mathrm{tr}$ of 0.27 (4). Note that $\nu_\mathrm{tr}$ is expressed in units of $\nu_\mathrm{max}$. Error bars derived from the range of possible intersection point of the two fits, using the uncertainties of the fitted parameters. Right axis (orange) shows the initial (first 100 ns) lifetime of emission as a function of the initial density $\nu_\mathrm{max}$. The power at which $\nu_\mathrm{max}$ = $\nu_\mathrm{cr}$ is indicated by the dashed line, and coincides with the point where a slower initial lifetime is first measured.
{\bf e.} Schematic of quadrupolar excitons at 1/3 filling. Moire unit cell is indicated in purple dashed lines, and the hexagonal full lines indicate the effective electronic lattice, with each electron site at the center of a hexagon. At 1/3 filling, assuming a homogeneous exciton distribution, every empty site is adjacent to at least one quadrupolar exciton. When an additional exciton is added, it necessarily resides next to three neighbors and may form an attractive state. {\bf f.} Diagram of decay rates of excitons with static in-plane dipolar and quadrupolar moments. Excitons with a static dipolar moment form a triply-degenerate basis, due to the triangular symmetry of the H-stacked potential. The oscillating dipole moment of the static quadrupolar state ($\gamma_{\rm Q}$) is an equal superposition of the oscillating dipole moments of the static dipolar states ($\gamma_{\rm d}$), and is expected to have a 3x faster collective decay rate ($\gamma_{\rm Q} = 3\gamma_{\rm d}$). Purple dashed lines indicate moire lattice edges. {\bf g.} Model fits of power-dependent lifetime data (see Supplementary Note 4). 

\newpage

\begin{figure}
\includegraphics[width = 6in]{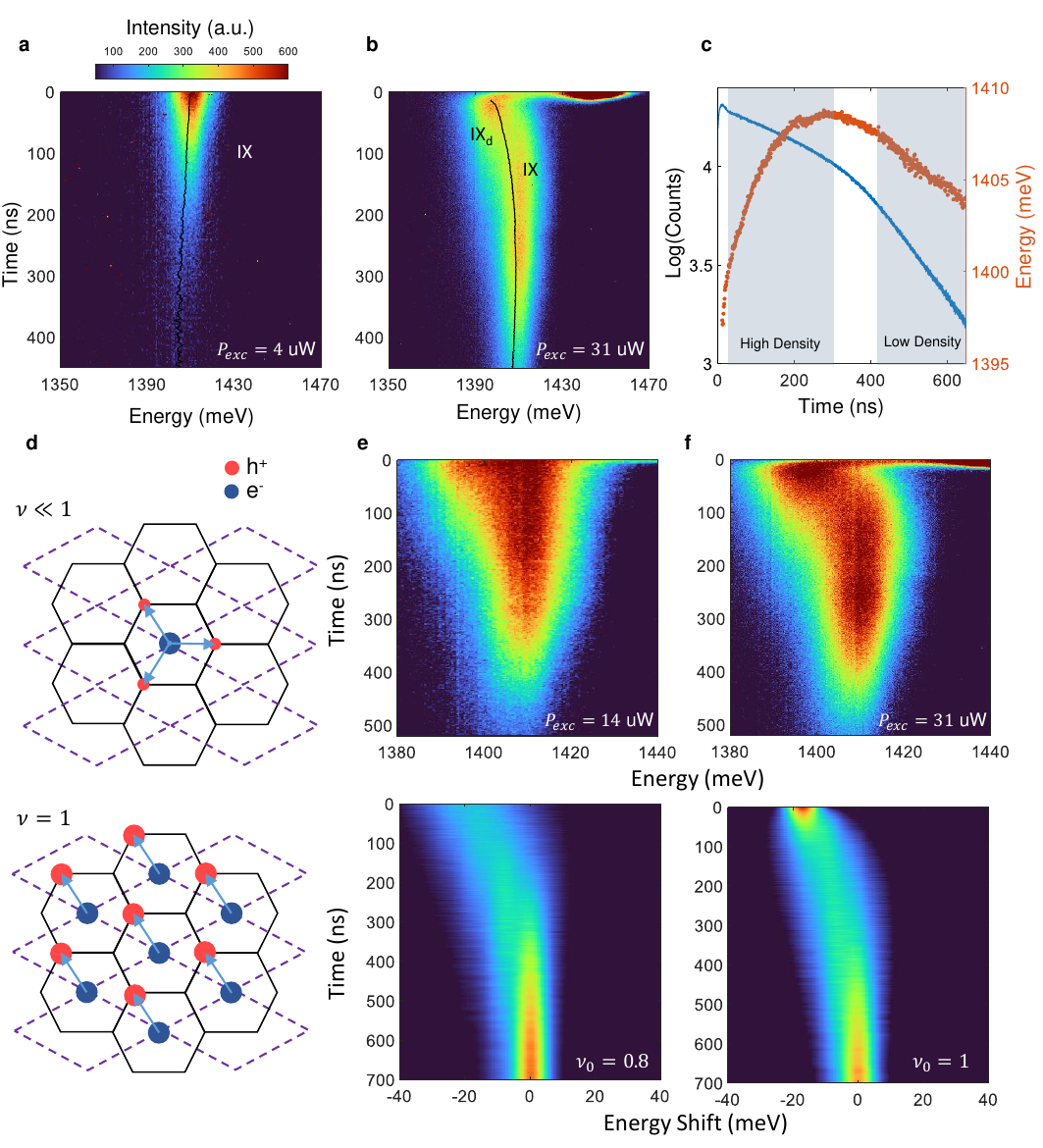}
\end{figure}

\noindent {\bf Figure 4: Modeling the density-dependent IX lifetime and energy dynamics in the H-stacked region.}
{\bf a.} Time-resolved IX spectrum at $\nu_\mathrm{max} =$ 0.4, using a 640nm pulsed laser. Black line indicates fitted peak center.
{\bf b.} Time-resolved IX spectrum at $\nu_\mathrm{max} >$ 1, using a 640nm pulsed laser. Black line indicates fitted peak center. Colorbar shared with panel {\bf a}.
{\bf c.} Decay profile and fitted peak center position of the data in panel {\bf b}. The emission center is initially red-shifted and blue-shifts during the period where the decay is slow.
{\bf d.} Diagram of transition from quadrupolar, low density exciton to an ferroelectric phase at unit filling.
{\bf e.} Comparison of measured (top) time-dependent PL at $P_\mathrm{exc}$ = 14 uW to simulation (bottom). Simulation panels show intensity-normalized spectra to emphasize the emission energy. The details of the simulation are given in Supplementary Note 3. Colorbar shared with panel {\bf a}.
{\bf f.} Comparison of measured (top) time-dependent PL at $P_\mathrm{exc}$ = 31 uW to simulation (bottom). Colorbar shared with panel {\bf a}.

\end{document}